\documentclass[a4paper,10pt,]{article}

\usepackage{authblk}

\newcount\eqnumber
\def\neweq{{\rm{(\the\eqnumber)}}\global\advance\eqnumber by 1}
\def\eqdef#1{\eqno\xdef#1{\the\eqnumber}\neweq}
\def\newaeq{{\rm{(\the\eqnumber a)}}\global\advance\eqnumber by 1}
\def\eqdaf#1{\eqno\xdef#1{\the\eqnumber}\newaeq}
\def\eqdisp#1{\xdef#1{\the\eqnumber}\neweq}
\def\eqdasp#1{\xdef#1{\the\eqnumber}\newaeq}

\newcount\refnumber
\def\newref{{\the\refnumber}\global\advance\refnumber by 1}
\def\refdef#1{{\xdef#1{\the\refnumber}}\newref}

\title{Constructing discrete Painlev\'e equations: from E$_8^{(1)}$ to A$_1^{(1)}$ and back}

\author[1]{A. Ramani and B. Grammaticos} 
\affil[1]{ IMNC,  CNRS, Universit\'e Paris-Diderot, Universit\'e Paris-Sud, Universit\'e Paris-Saclay, 91405 Orsay, France\medskip}

\author[2]{R. Willox}
\affil[2]
{\sl Graduate School of Mathematical Sciences, the University of Tokyo, 3-8-1 Komaba, Meguro-ku, 153-8914 Tokyo, Japan \medskip}

\author[3]{T. Tamizhmani}
\affil[3]{\sl SAS, Vellore Institute of Technology, Vellore - 632014, Tamil Nadu, India}

\begin{document}

\maketitle


\begin{abstract}\noindent
The `restoration method' is a novel method we recently introduced for systematically deriving discrete Painlev\' e equations. In this method we start from a given Painlev\' e equation, typically with E$_8^{(1)}$ symmetry, obtain its autonomous limit and construct all possible QRT-canonical forms of mappings that are equivalent to it by homographic transformations. Discrete Painlev\'e equations are then obtained by deautonomising the various mappings thus obtained. We apply the restoration method to two challenging examples, one of which does not lead to a QRT mapping at the autonomous limit but we verify that even in that case our method is indeed still applicable. For one of the equations we derive we also show how, starting from a form where the independent variable advances one step at a time, we can obtain versions that correspond to  multiple-step evolutions. 

\end{abstract}


\section{ Introduction}

The classification of discrete Painlev\'e equations, based on the pioneering work of Sakai [\refdef\sakai], not only brought much needed order to the domain but also showed that there exists a third type of discrete Painlev\'e equation, the elliptic type, besides the additive and multiplicative ones: elliptic equations are non-autonomous discrete systems where the independent variable enters through the argument of elliptic functions. The derivation of elliptic discrete Painlev\'e equations based on the deautonomisation procedure, however, necessitated the introduction of a new ansatz, which we proposed in [\refdef\ellip] in collaboration with Y. Ohta and which we dubbed trihomographic. The latter has the form
$${x_{n+1}-a\over x_{n+1}-b}{x_{n-1}-c\over x_{n-1}-d}{x_{n}-e\over x_{n}-f}=g,\eqdef\zena$$
and was directly inspired by the Miura transformation [\refdef\eight] in E$_8^{(1)}$ space. With this ansatz it was indeed possible to derive the first concrete examples of elliptic discrete Painlev\'e equations. However, the form (\zena) is not only tailored to the elliptic type but obviously also encompasses the additive and multiplicative equations as well [\refdef\multi]. We have indeed for the additive equations the form
$${x_{n+1}-a_n^2\over x_{n+1}-b_n^2}\ {x_{n-1}-c_n^2\over x_{n-1}-d_n^2}\ {x_{n}-e_n^2\over x_{n}-f_n^2}=1,\eqdef\zdyo$$
and for the multiplicative ones,
$${x_{n+1}-\sinh^2a_n\over x_{n+1}-\sinh^2b_n}\ {x_{n-1}-\sinh^2c_n\over x_{n-1}-\sinh^2d_n}\ {x_{n}-\sinh^2e_n\over x_{n}-\sinh^2f_n}=1,\eqdef\ztri$$
where $a_n,b_n,\cdots,f_n$ are specific linear functions of the independent variable $n$ (the same for (\zdyo) and (\ztri), and in fact for the elliptic equation that follows). The corresponding trihomographic form for the elliptic equations is
$${x_{n+1}-{\rm sn}^{2}a_n\over x_{n+1}-{\rm sn}^{2}b_n}\ {x_{n-1}-{\rm sn}^{2}c_n\over x_{n-1}-{\rm sn}^{2}d_n}\ {x_{n}-{\rm sn}^{2}e_n\over  x_{n}-{\rm sn}^{2}f_n}
= {\theta_0^2(b_n)\over \theta_0^2(a_n)}\ {\theta_0^2(d_n)\over \theta_0^2(c_n)}\ {\theta_0^2(f_n)\over  \theta_0^2(e_n)}.\eqdef\ztes$$ 
The trihomographic form is not limited to the representations of equations associated to the E$_8^{(1)}$ affine Weyl group, as we have shown in [\refdef\jmptri]. In fact {\sl all} discrete Painlev\'e equations can be cast into a trihomographic form. On the other hand, a single trihomographic equation is not equivalent to the most general discrete Painlev\'e equation. Let us take the example of the additive E$_8^{(1)}$-associated equation. The form of the latter is
$${(x_n-x_{n+1}+(z_n+z_{n+1})^2)(x_n-x_{n-1}+(z_n+z_{n-1})^2)+4x_n(z_n+z_{n+1})(z_n+z_{n-1})\over 
(z_n+z_{n-1})(x_n-x_{n+1}+(z_n+z_{n+1})^2)+(z_n+z_{n+1})(x_n-x_{n-1}+(z_n+z_{n-1})^2)}
=R(x_n),\eqdef\zpen$$
where in the general case $R$ is a ratio of two specific polynomials of $x$, quartic in the numerator and cubic in the denominator. Working with the trihomographic form (\zdyo) with 
$$\displaylines{a_n=z_n+z_{n-1}+k_n,\ b_n=z_n+z_{n-1}-k_n,\ c_n=z_n+z_{n+1}+k_n,\ d_n=z_n+z_{n+1}-k_n, \hfill\cr\hfill e_n=2z_n+z_{n-1}+z_{n+1}-k_n,\ f_n=2z_n+z_{n-1}+z_{n+1}+k_n,\quad\eqdisp\zhex\cr}$$
we can show formal equivalence of the latter to (\zpen) provided that
$$R(x_n)={x_n-k_n^2\over 2z_n+z_{n-1}+z_{n+1}}+2z_n+z_{n-1}+z_{n+1}.\eqdef\zhep$$
As explained in [\jmptri], in order to obtain the most general right-hand side in (\zpen) one has to couple four trihomographic equations together. Still, working with a single trihomographic equation can lead to very rich results, as we showed in [\refdef\deight].
Moreover, a glimpse at equations (\zdyo), (\ztri) and (\ztes) suffices to convince oneself that there is no need to address the construction of the multiplicative and elliptic systems separately. Once the form of the additive equation is established, i.e. when the values of $z_n$ and $k_n$ in (\zhex) have been obtained, one can transcribe them directly to the multiplicative or elliptic cases.

In this paper we shall use the trihomographic representation in order to construct discrete Painlev\'e equations using the deconstruction/restoration method we introduced in [\refdef\resto]. We can illustrate the workings of this method through some very simple example. Consider the autonomous limit of the general trihomographic discrete Painlev\'e equation, which has the form:
$${x_{n+1}-a\over x_{n+1}-b}{x_{n-1}-a\over x_{n-1}-b}={x_{n}-c\over x_{n}-d}.\eqdef\zoct$$
Next, introduce the homographic transformation 
$$X_n=f\, {x_{n}-a\over x_{n}-b},\eqdef\zenn$$
and rewrite (\zoct) in the simple form 
$$X_{n+1}X_{n-1}={AX_{n}-B\over X_{n}-1},\eqdef\zdek$$
where $f=(b-d)/(a-d)$ and $A= f^2 (b-c)/(b-d)$, $B= f^2(a-c)/(a-d)$. This concludes the {\sl deconstruction} part of the procedure and we call (\zdek) the {\sl remnant} equation. Since the latter is a mapping of QRT [\refdef\qrt] type we can easily obtain its invariant. In the present case we find
$$K={X_n^2X_{n-1}^2-(A+1)X_nX_{n-1}(X_n+X_{n-1})+A(X_n^2+X_{n-1}^2)-(A^2+B)(X_n+X_{n-1})+AB\over X_nX_{n-1}}.\eqdef\dena$$
The {\sl restoration} phase then starts by introducing a new homographic transformation
$$X_n={p x_{n}+q\over r  x_{n}+s},\eqdef\ddyo$$
in which one chooses the values of $p,q,r,s$ ($p s\neq q r$) so as to bring the invariant ${\cal K}=(K+\mu)^{-1}$, with $\mu$ a properly chosen constant, to one of the canonical forms already catalogued [\refdef\auton,\refdef\canon] for the QRT mappings. The deautonomisation of the QRT mapping thus obtained leads to a discrete Painlev\'e equation, different from the initial one if the homography (\ddyo) is chosen to be different from (\zenn). A systematic search for such homographies then yields an entire cascade of discrete Painlev\'e equations, with various types of symmetries, starting from just a single one.
 
 Two different discrete Painlev\'e equations, obtained in [\deight], will be considered in this paper. Both are expressed in (in QRT parlance) `asymmetric'  trihomographic form. (Note that while in this introduction, for the sake of simplicity, our presentation was based on symmetric trihomographic forms, our arguments apply with minimal changes to the asymmetric case as well). The general form of an asymmetric trihomographic system is
$${x_{n+1}-a_n\over x_{n+1}-b_n}{x_{n}-c_n\over x_{n}-d_n}{y_{n}-e_n\over y_{n}-f_n}=1,\eqdaf\dtri$$
$${y_{n}-g_n\over y_{n}-h_n}{y_{n-1}-p_n\over y_{n-1}-q_n}{x_{n}-r_n\over x_{n}-s_n}=1,\eqno(\dtri b)$$
where the parameters are given by:
$$\displaylines{a_n=(\zeta_n+z_n+k_n)^2,\ b_n=(\zeta_n+z_n-k_n)^2,\ c_n=(\zeta_n+z_{n+1}+k_n)^2,\ d_n=(\zeta_n+z_{n+1}-k_n)^2, \hfill\cr\hfill e_n=(2\zeta_n+z_n+z_{n+1}-k_n)^2,\ f_n=(2\zeta_n+z_n+z_{n+1}+k_n)^2,\quad\eqdasp\dtes\cr}$$
\vskip-1.3cm
$$\displaylines{g_n=(\zeta_{n-1}+z_n+\kappa_n)^2,\ h_n=(\zeta_{n-1}+z_n-\kappa_n)^2,\ p_n=(\zeta_n+z_{n}+\kappa_n)^2,\ q_n=(\zeta_n+z_{n}-\kappa_n)^2, \hfill\cr\hfill r_n=(\zeta_n+2z_n+\zeta_{n-1}-\kappa_n)^2,\ s_n=(\zeta_n+2z_n+\zeta_{n-1}+\kappa_n)^2.\quad(\dtes b)\cr}$$
Before giving the precise $n$-dependence for the two equations we will study, let us introduce some useful notation. The independent variable will enter  through a secular term of the form $t_n=\alpha n+\beta$ and two types of periodic functions:  $\phi_m(n)$ and $\chi_{2m}(n)$. The first one has period $m$, i.e. $\phi_m(n+m)=\phi_m(n)$, and is given by
 $$ \phi_m(n)=\sum_{\ell=1}^{m-1} \epsilon_{\ell}^{(m)} \exp\left({2i\pi \ell n\over m}\right).\eqdef\dpen$$
The summation starts at 1 instead of 0 and thus $\phi_m$ introduces $(m-1)$ parameters. 
The second periodic function $\chi_{2m}$ obeys the equation $\chi_{2m}(n+m)+\chi_{2m}(n)=0$. It has period $2m$, while involving only $m$ parameters, and can be expressed in terms of roots of unity as
$$ \chi_{2m}(n)=\sum_{\ell=1}^{m} \eta_{\ell}^{(m)} \exp\left({i\pi(2\ell-1)n\over m}\right).\eqdef\dhex$$
The first of the equations we consider is case VII in the numbering system introduced in [\multi] where we presented the derivation of additive E$_8^{(1)}$-associated trihomographic equations. The parameters in this equation are given by: 
$$u_n=t_n+\phi_4(n)+\phi_5(n),\ z_n+\zeta_n=u_{n+1}-2u_n,\ z_{n+1}+\zeta_n=u_{n-1}-2u_n,\ k_n=u_n,\ \kappa_n=u_{n+1}+u_n+u_{n-1}+u_{n-2}.$$
For the second equation, case X in [\multi],  the parameters are:
$$u_n=t_n+\phi_4(n),\ z_n+\zeta_n=u_{n}-\gamma,\ z_{n+1}+\zeta_n=u_{n}+\gamma,\ k_n=u_{n+1}-u_n+u_{n-1}+\phi_2(n),\ \kappa_n=\delta+\widetilde{\phi}_2(n),$$
where $\phi_2(n)$ and $\widetilde{\phi}_2(n)$ are two independent functions of period 2.  A small remark is in order here. In [\multi] we also identified an equation, case XI, with parameters 
$$u_n=t_n+\phi_4(n),\ z_n+\zeta_n=u_{n}-\gamma,\ z_{n+1}+\zeta_n=u_{n}+\gamma,\ k_n=u_{n+1}-u_n+u_{n-1}+\phi_2(n),\ \kappa_n={\chi}_4(n).$$
However, it turns out that the latter is nothing but a rewriting of case X. Indeed, it suffices to exchange numerator and denominator of two successive instances out of four in case X. This is tantamount to changing the sign of $\kappa_n$ two times out of four and, thanks to this inversion, we obtain a $\chi_4(n)$ when starting from $\delta+\phi_2(n)$ and vice versa. Note that, while cases X and XI are perfectly equivalent when non-autonomous, they do not give the same autonomous limit since in the case of X the parameter $\delta$ survives at the autonomous limit, while for XI we have $\kappa_n\equiv0$.
In the following sections we shall apply the deconstruction/restoration procedure to cases VII, X, and XI and derive the discrete Painlev\'e equations that can be obtained with this method. 

Another interesting construction of discrete Painlev\'e equations, which we introduced in [\refdef\ouranton] and [\refdef\mittami] is one based on multistep evolutions. Depending on the equation at hand it may be possible, for example, to skip one out of two indices and obtain a mapping relating variables at a distance of two. We can easily illustrate this construction on the formal remnant equation (\zdek). Using (\zdek) to eliminate $X_n$ in terms of $X_{n+1}$ and $X_{n-1}$, we find the invariant
$$K={(1-A)X_{n+1}X_{n-1}(X_{n+1}+X_{n-1})+(A^2-A-B-1)X_{n+1}X_{n-1}+(A^2+B)(X_{n+1}+X_{n-1})-A^3-B\over (X_{n+1}X_{n-1}-A)(X_{n+1}X_{n-1}-B)}.\eqdef\dhep$$
Using (\dhep) we can obtain a double-step evolution. We find
$$\left({X_nX_{n+2}-B\over X_nX_{n+2}-A}\right)\left({X_nX_{n-2}-B\over X_nX_{n-2}-A}\right)={AX_n-B\over X_n-1},\eqdef\doct$$
which can be put into canonical form by the appropriate scaling of $X$ and then deautonomised to a discrete Painlev\'e equation. 
Higher, multistep, evolutions may also be possible and in what follows we shall present examples thereof.

\section{ The trihomographic equation with periods 4 and 5}

The first equation we are going to study is the case VII we referred to in the introduction. It is given by the asymmetric trihomographic form (\dtri) with parameters $u_n=t_n+\phi_4(n)+\phi_5(n)$, $z_n+\zeta_n=u_{n+1}-2u_n$, $z_{n+1}+\zeta_n=u_{n-1}-2u_n$, $k_n=u_n$ and $\kappa_n=u_{n+1}+u_n+u_{n-1}+u_{n-2}$. At the autonomous limit we neglect the secular and periodic dependences and, taking $\beta=1$, we obtain the mapping
$${x_{n+1}-4\over x_{n+1}}{x_{n}-4\over x_n}{y_{n}-1\over y_{n}-9}=1,\eqdaf\denn$$
$${y_{n}-9\over y_{n}-25}{y_{n-1}-9\over y_{n-1}-25}{x_{n}-36\over x_{n}-4}=1.\eqno(\denn b)$$
Next we introduce the homographic transformations
$$X_n={8\over9}{x_n\over x_n-4},\qquad Y_n={1\over3}{y_n-25\over y_n-9},\eqdef\ddek$$
and obtain the remnant system
$$X_{n+1}X_n=A(1-Y_n),\eqdaf\vena$$
$$Y_nY_{n-1}=1-X_n,\eqno(\vena b)$$
with $A=32/27$. However, in what follows, in order to have full freedom we shall consider $A$ to be a free, non-zero, parameter. The invariant of (\vena) can be easily obtained: 
$$K={X_nY_n(X_n+Y_n)-X_n^2-AY_n^2+(A+1)X_n+2AY_n-A\over X_nY_n}.\eqdef\vdyo$$
Equation (\vena) can be deautonomised with $A_n=\alpha n+\beta+\phi_4(n)$, resulting in a discrete Painlev\'e equation associated to the affine Weyl group A$_4^{(1)}$ [\refdef\qasym]. 

Given the form of (\vena) it is clear that one can eliminate either of the variables and obtain an equation for $X$ or $Y$ alone (and, obviously, this can be done on the deautonomised forms as well). Eliminating $X$ we find for $Y$ the equation
$$(Y_nY_{n+1}-1)(Y_nY_{n-1}-1)=A_n(1-Y_n),\eqdef\vtri$$
where $A_n$ is the same as in the previous paragraph.
Similarly, eliminating $Y$ we find for $X$ the equation
$$(X_{n+1}X_n-A_n)(X_{n-1}X_n-A_{n-1})=A_nA_{n-1}(1-X_n), \eqdef\vtes$$
which is not precisely in canonical form (but which can be cast into one by the appropriate scaling of $X$). Interestingly, given the form of (\vtes) it is possible to obtain a double step evolution
$$X_{n+2}X_{n-2}={A_{n+1}A_{n-2}\over A_{n}A_{n-1}}{(X_n-A_n)(X_n-A_{n-1})\over1-X_n},\eqdef\vpen$$
a discrete Painlev\'e equation equation already obtained in [\qasym], equation 11.

Multistep evolutions can be obtained also for the initial, asymmetric, system. 
The equation corresponding to a triple-step evolution (or, in fact, a triple half-step evolution if we consider that each of the equations of (\vena) is a half-step one, a full step being obtained by taking both equations) is:
$$(X_{n}Y_{n+1}-A_n)(X_{n+3}Y_{n+1}-A_{n+2})={A_nA_{n+2}\over A_{n+1}}\Big(1-Y_{n+1}\Big)\eqdaf\vhex$$
$$(X_{n}Y_{n-2}-A_{n-1})(X_{n}Y_{n+1}-A_n)={(X_n-A_n)(X_n-A_{n-1})\over1-X_n},\eqno(\vhex b)$$
an equation already derived in [\qasym], equation 127.
A quintuple-step evolution is also possible: 
$$\left({A_{n+5}Y_{n+3}-X_{n+6}\over A_{n+5}Y_{n+3}-A_{n+4}X_{n+6}}\right)\left({A_{n+1}Y_{n+3}-X_{n+1}\over A_{n+1}Y_{n+3}-A_{n+2}X_{n+1}}\right)={A_{n+3}\over A_{n+2}A_{n+4}}\Big(1-Y_{n+3}\Big),\eqdaf\vhep$$
$$\left({A_{n+1}Y_{n+3}-X_{n+1}\over A_{n+1}Y_{n+3}-A_{n+2}X_{n+1}}\right)\left({A_nY_{n-2}-X_{n+1}\over A_nY_{n-2}-A_{n-1}X_{n+1}}\right)={1\over A_nA_{n+1}}{(X_{n+1}-A_n)(X_{n+1}-A_{n+1})\over1-X_{n+1}}.\eqno(\vhep b)$$
which, to the authors' best knowledge, is a discrete Painlev\'e equation which has not been previously derived.

Before proceeding to the restoration starting from (\vena) it is interesting to consider the $x$- or $y$-only mappings obtained from (\denn). Eliminating $x$ we find the equation
$${(y_n-y_{n+1}+4)(y_n-y_{n-1}+4)+16y_n\over2y_n-y_{n+1}-y_{n-1}+8}={y_n^2+2y_n-75\over2y_n-20}.\eqdef\voct$$
We could, of course, have performed the elimination of $x$ at the level of the full, i.e. including secular and periodic dependence, VII system. In this case we would have obtained directly an additive Painlev\'e equation related to E$_8^{(1)}$, which turns out to be among those derived in [\refdef\addit] (equation 4.5.2). On the other hand, the elimination of $y$ in (\denn) leads to an equation involving only $x$, which, as it turns out, can be cast into the trihomographic form:
$${x_{n+1}-16\over x_{n+1}}{x_{n-1}-16\over x_{n-1}}{x_{n}-4\over x_{n}-36}=1.\eqdef\venn$$
As in the case of (\voct), we could have performed the elimination on the full case VII obtaining a trihomographic form. The latter is precisely the equation dubbed Case I in [\multi], and which was indeed first derived in [\ellip], where its elliptic extension was also presented. 

Since (\venn) is trihomographic it is possible to obtain a double-step evolution for it. Working in the autonomous case we find the mapping
$${(x_n-x_{n+2}+16)(x_n-x_{n-2}+16)+64x_n\over2x_n-x_{n+2}-x_{n-2}+32}={x_n^2+92x_n+192\over2x_n+28}.\eqdef\vdek$$
which, when deautonomised becomes equation 4.2.1 of [\addit].
Introducing the homographic transformation
$$X=-3{x_{n}-16\over x_{n}},\eqdef\tena$$
we obtain from (\venn) the remnant equation
$$X_{n+1}X_{n-1}={X_n-B\over X_n-1},\eqdef\tdyo$$
with $B=-5/27$ ($=1-A$). This equation is of QRT-type and its invariant is
$$K={X_n^2X_{n-1}^2-2X_nX_{n-1}(X_n+X_{n-1})+X_n^2+X_{n-1}^2-(B+1)(X_n+X_{n-1})+B\over X_nX_{n-1}}.\eqdef\ttri$$
The deautonomisation of (\tdyo) leads to $B_n=\alpha n+\beta+\phi_5(n)$, a discrete Painlev\'e equation associated to D$_5^{(1)}$, first obtained in [\qasym] (equation A1ciii). Given the form of (\tdyo) we can obtain readily a double step evolution. We find thus 
$$\left({X_nX_{n+2}-B_{n+1}\over X_nX_{n+2}-1}\right)\left({X_nX_{n-2}-B_{n+1}\over X_nX_{n-2}-1}\right)={X_n-B_n\over X_n-1},\eqdef\ttes$$
a discrete Painlev\'e equation first derived in [\refdef\sequel].

We now proceed to the restoration phase starting from the remnant equation (\vena). We introduce the transformation
$$X_n={ax_n+b\over cx_n+d},\qquad Y_n={py_n+q\over ry_n+s},\eqdef\tpen$$
and search for the possible choices of the $a, b,\cdots,r, s$ that can bring the invariant (\vdyo) to one of the QRT canonical forms. Clearly, if we use transformation (\ddek) we go back from (\vena) to the initial system (\denn), and upon deautonomisation to the additive E$_8^{(1)}$ associated discrete Painlev\'e equation we already identified as case VII in [\multi]. 

Introducing the homographies, for $z^8\neq1$ lest they become trivial, 
$$X_n={(z^2+1/z^2)(z+1/z)^2\over((z+1/z)^2-1)^2}\,{x_n+2\over x_n+z^2+1/z^2},\qquad Y_n={1\over (z+1/z)^2-1}\,{y_n+z^5+1/z^5\over y_n+z^3+1/z^3},\eqdef\thex$$
we obtain the invariant
$${\cal K}={x_n^2-x_ny_n(z+1/z)+y_n^2+(z-1/z)^2\over(x_n+2)(x_n+z^2+1/z^2)(y_n+z^5+1/z^5)(y_n+z^3+1/z^3)},\eqdef\thep$$
where $z$ and $A$ are related through $A=(z^2+1/z^2)(z+1/z)^4/((z+1/z)^2-1)^3$ and the quantity $\mu$ is given by 
$\mu=(3(z^4+1/z^4)+8(z^2+1/z^2)+8)/((z+1/z)^2-1)^2$. This invariant leads to a multiplicative mapping which, upon deautonomisation, produces a discrete Painlev\'e equation associated to E$_8^{(1)}$. (As mentioned in the introduction, in [\multi] we have explained how, once the additive E$_8^{(1)}$-associated discrete Painlev\'e equation is obtained one can construct in an elementary way the multiplicative and elliptic analogues).

At this point a remark is in order. The multiplicative mapping obtained in the previous paragraph was derived by assuming that the parameter $A$, appearing in the remnant equation, was a free one and not fixed to the value 32/27 that arises in the deconstruction of (\denn). Does this mean that there is no multiplicative E$_8^{(1)}$-type mapping which, at the autonomous limit, corresponds to mapping (\denn) with $A=32/27$? It turns out that in this case restoration towards a multiplicative mapping is possible even when we work with $A=32/27$. For this it suffices to see that the parameter $A$ takes this particular value not only when $z^2=1$, which makes (\thex) pathological and leads back to (\vena), but also when $z+1/z=\pm\sqrt{2/5}$. Note that because of obvious symmetries, all four solutions to this relation yield the same value for the constant $\mu$: $\mu=-26/3$. We find in this case the homographies
$$X_n=-{80\over9}\,{x_n+2\over 5x_n-8},\qquad Y_n={\mp25\sqrt{10}y_n+158\over \pm15\sqrt{10}y_n+78}.\eqdef\toct$$
So, while working with the fixed value of parameters is not a real limitation, freeing them allows for a more convenient form of the restored mapping, one that would be easily amenable to deautonomisation. 

It turns out that no restoration to equations associated with E$_7^{(1)}$ or E$_6^{(1)}$ is possible. The only possible restoration is one leading to D$_5^{(1)}$-associated equations.  We introduce the homographies
$$X_n={Ax_n\over x_n-1},\qquad Y_n={y_n-1+A\over y_n-1},\eqdef\tenn$$
leading to the invariant
$${\cal K}={x_ny_n-1\over x_n(x_n-1)(y_n-1)(y_n-1+A)},\eqdef\tdek$$
with $\mu=-2-A$. The corresponding mapping is 
$$(x_{n+1}y_n-1)(x_ny_n-1)=1-y_n\eqdaf\qena$$
$$(x_ny_{n-1}-1)(x_ny_n-1)=x_n^2(1-A)-x_n(2-A)+1.\eqno(\qena b)$$
Given the form of (\qena) we can eliminate $y$ and obtain an equation for $x$ alone. We obtain thus the equation, 
$$x_{n+1}x_{n-1}={x_n-1\over Bx_n-1},\eqdef\qdyo$$
with $B=1-A$ and by inverting $x$ we recover exactly equation (\tena). Similarly we can eliminate $x$ and obtain an equation for $y$ alone:
$$(y_{n+1}y_n-B)(y_{n-1}y_n-B)=(y_n-B)^2(y_n-1),\eqdef\qtri$$
the deautonomisation of which was given in [\qasym], equation 65. 

\section{ The trihomographic equation with periods 2, 2, and 4}

We turn now to the equation referred to in the introduction as case X. It has the form of a trihomographic mapping (\dtri) with parameters $u_n=t_n+\phi_4(n), \ z_n+\zeta_n=u_{n}-\gamma, \ z_{n+1}+\zeta_n=u_{n}+\gamma, \ k_n=u_{n+1}-u_n+u_{n-1}+\phi_2(n), \kappa_n=\delta+\widetilde{\phi}_2(n)$. At the autonomous limit we neglect the secular and periodic dependence and, taking $\beta=1$, obtain the mapping
$${x_{n+1}-(2-\gamma)^2\over x_{n+1}-\gamma^2}{x_{n}-(2+\gamma)^2\over x_n-\gamma^2}{y_{n}-1\over y_{n}-9}=1,\eqdaf\qtes$$
$${y_{n}-(1+\gamma+\delta)^2\over y_{n}-(1+\gamma-\delta)^2}{y_{n-1}-(1-\gamma+\delta)^2\over y_{n-1}-(1-\gamma-\delta)^2}{x_{n}-(2-\delta)^2\over x_{n}-(2+\delta)^2}=1.\eqno(\qtes b)$$
Due to the presence of $\gamma$ this mapping is not of QRT type. This can be easily assessed when one considers the invariant of (\qtes) together with the conservation condition. We have indeed an `invariant'
$$K(x_n, y_n;\gamma)={(1-\gamma)(x_n-\delta^2)(y_n-1)\big(x_ny_n(1-\gamma)+x_n(2\gamma^2+\gamma-2\delta^2-1)+y_n(2\gamma^2+(\gamma+1)\delta^2-4)+\theta\big)\over (x_n-y_n)^2 - 2(x_n+y_n)(\gamma - 1)^2 +(\gamma - 1)^4},
\eqdef\qpen
$$
where $\theta=-2\gamma^4+2\gamma^2(\delta^2-1)-\delta^2(\gamma+1)+4$.
However the conservation law, instead of the usual QRT one $K(x_n, y_{n-1})=K(x_n, y_n)=K(x_{n+1}, y_n)$, is now 
$$K(x_n, y_{n-1};-\gamma)=K(x_n, y_n;\gamma)=K(x_{n+1}, y_n;-\gamma).\eqdef\qhex$$ 
Since (\qtes) is not a QRT mapping one cannot implement the restoration procedure directly. (Of course, when $\gamma=0$ one recovers a QRT mapping and the method proceeds as normal ; we will examine this special case at the end of this section.) In [\resto], however, we have suggested a way to tackle non-QRT remnant mappings and in the following we will show that, using a similar procedure, we can indeed perform the restoration for the case $\gamma\ne0$ here as well.

We start by considering the equations obtained from (\qtes) when we eliminate one of the two variables. Eliminating $y$ we obtain, for $x$ alone, the mapping
$${(x_n-x_{n+1}+4)(x_n-x_{n-1}+4)+16x_n\over2x_n-x_{n+1}-x_{n-1}+8}={x_n^2-x_n(\gamma^2+\delta^2-24)+(\gamma^2-4)(\delta^2-4)\over2x_n-\gamma^2-\delta^2+8},\eqdef\qhep$$
which {\sl is} of QRT-type. Its deautonomisation of (\qhep) was presented in [\addit], equation 4.1.
The invariant corresponding to (\qhep) is
$$K(x)={x_n^2x_{n-1}^2-x_nx_{n-1}(x_n+x_{n-1})(\gamma^2+\delta^2)+x_nx_{n-1}(\gamma^4+4\delta^2\gamma^2+\delta^4)-(x_n+x_{n-1})\gamma^2\delta^2(\gamma^2+\delta^2-8)+\theta\over (x_n-x_{n-1})^2 - 8(x_n+x_{n-1})+16},\eqdef\qoct$$
where $\theta=\gamma^2\delta^2(\gamma^2+\delta^2-16)$.
The usual deconstruction/restoration procedure consists in simplifying (\qhep) through a homographic transformation, obtaining the remnant mapping. However, this is not a mandatory step. We can perfectly well start from the initial mapping (\qhep) dispensing with the deconstruction part of the procedure. Of course, if one wishes to work with the initial mapping, it would be better in that case to start with the multiplicative E$_8^{(1)}$ one rather than the additive. However even in the latter case one obtains the same restoration results for equations with E$_7^{(1)}$ or D$_5^{(1)}$ symmetries.

We introduce the transformation
$$x_n={aX_n+b\over X_n+c},\eqdef\qenn$$
and seek the possible QRT-canonical forms of the invariant ${\cal K}=(K(x)+\mu)^{-1}$. The first canonical form is obtained, for $\mu=\gamma^2\delta^2$, with the homography
$$x_n={\gamma^2X_n-\delta^2\over X_n-1}.\eqdef\dqek$$
It has the form
$$X_{n+1}X_{n-1}={(X_n-A)(X_n-B)\over(1-CX_n)(1-DX_n)},\eqdef\pena$$
where $A,B,C,D$ are expressed in terms of the $\gamma,\delta$. The deautonomisation of this equation was presented in [\refdef\dps]. It was shown,  by Jimbo and Sakai [\refdef\jimbo], to be a discrete analogue of the Painlev\'e VI equation and is associated to the affine Weyl group D$_5^{(1)}$.

Eliminating $x$ we find for $y$ the mapping, which is again of QRT type,
$${(y_n-y_{n+1}+4)(y_n-y_{n-1}+4)+16y_n\over2y_n-y_{n+1}-y_{n-1}+8}={y_n^3-y_n^2(2\gamma^2+2\delta^2-25)+y_n((\gamma^2-\delta^2)^2-24(\gamma^2+\delta^2)+35)+3\eta\over2y_n^2-y_n(3\gamma^2+3\delta^2-10)+(\gamma^2-\delta^2)^2-5(\gamma^2+\delta^2)+4},\eqdef\pdyo$$
with $\eta=((\gamma+\delta)^2-1)((\gamma-\delta)^2-1)$, the non-autonomous form of which was obtained in [\addit], equation 5.1.1.
The invariant of (\pdyo) is
$$K(y)={y_n^2y_{n-1}^2-y_ny_{n-1}(y_n+y_{n-1})(\gamma^2+\delta^2)+y_ny_{n-1}((\gamma^2-\delta^2)^2+6(\gamma^2+\delta^2)-6)-(y_n+y_{n-1})\lambda+\theta
\over (y_n-y_{n-1})^2 - 8(y_n+y_{n-1})+16},\eqdef\ptri$$
where $\lambda=((\gamma^2-\delta^2)^2-5(\gamma^2+\delta^2)+4)$ and $\theta=((\gamma^2-\delta^2)^2-14(\gamma^2+\delta^2)+13)$.
We introduce the transformation
$$y_n={pY_n+q\over Y_n+r},\eqdef\ptes$$
and look for the QRT-canonical forms of the invariant. We find that with the same value of $\mu$ as for the $x$ equation, i.e. 
$\mu=\gamma^2\delta^2$ and the homography
$$y_n={pY_n+(2-p)(2-(\delta-\gamma)^2)+2\over Y_n+2p+2-(\delta-\gamma)^2}.\eqdef\ppen$$
where $p$ obeys the constraint $p^2(\delta\gamma+1)-2p(\delta^2+3\delta\gamma+\gamma^2+1)+\delta^4-2\delta^2\gamma^2+\gamma^4-2(\delta^2+\gamma^2)+1=0$,
we obtain an equation of the form
$$(Y_{n+1}Y_n-1)(Y_nY_{n-1}-1)={(1-PY_n)(1-QY_n)(1-RY_n)(1-SY_n)\over 1-FY_n}.\eqdef\phex$$
The deautonomisation of this mapping was first obtained in [\refdef\mitjimbo] leading to an equation associated with the affine Weyl group D$_5^{(1)}$. In [\refdef\afb] we showed that the non-autonomous form of (\phex) is a discrete analogue of the Ablowitz-Fokas-Bureau equation, which is obtained from Painlev\'e VI by a Miura transformation. The $q$-difference analogue of this Miura transformation was given in [\refdef\ohta]. This Miura transformation, considered as a discrete Painlev\'e equation, is what one would expect to obtain from the restoration procedure, starting from (\qtes), had the application of the procedure been possible. 

While a restoration to an equation associated to E$_6^{(1)}$ is not possible, it turns out that we can obtain mappings that can be deautonomised to equations related to E$_7^{(1)}$. This canonical form is obtained with $\mu=(\delta^2+\gamma^2)^2/4-(\delta\pm\gamma)^2$. However, as the calculations are particularly voluminous no details can be presented here and we have to content ourselves with giving the final result. The equation for $X$ has the form
 $$\left({X_{n+1}X_n-Q_{n+1}Q_n\over X_{n+1}X_n-1}\right)\left({X_nX_{n-1}-Q_nQ_{n-1}\over X_nX_{n-1}-1}\right)={(X_n-A_n)(X_n-B_n)\over (X_n-R_n)(X_n-S_n)},\eqdef\phep$$
 where $\log Q_n=2\alpha n+2\beta+2n\phi_2(n)$, $\log R_n=-\alpha n-\beta+\phi_2(n)+\phi_4(n)+\eta$, $\log S_n=-\alpha n-\beta+\phi_2(n)-\phi_4(n)+\theta$, $\log A_n=3\alpha n+3\beta+\phi_2(n)+\phi_4(n+2)+\eta$, and $\log B_n=3\alpha n+3\beta+\phi_2(n)-\phi_4(n+2)+\theta$. Similarly for $Y$, we find, after deautonomisation the equation
 $$\left({Y_{n+1}Y_n-Q_{n+1}Q_n\over Y_{n+1}Y_n-1}\right)\left({Y_nY_{n-1}-Q_nQ_{n-1}\over Y_nY_{n-1}-1}\right)={(Y_n-A_n)(Y_n-BQ_n)(Y_n-Q_n/B)\over (Y_n-R_n)(Y_n-C)(Y_n-1/C)},\eqdef\poct$$
where $\log Q_n=\alpha n+\beta+\phi_4(n)$, $\log R_n=\phi_2(n)$, $A_n=2\alpha n+2\beta+\phi_2(n)$, and $B,C$ are constant. Both (\phep) and (\poct) were first derived in [\sequel], eqs. 2.16 and 2.9 respectively, where they were identified as discrete Painlev\'e equations associated to E$_7^{(1)}$. 

Additive analogues to equations (\phep) and (\poct) above do exist, provided the constraint $\pm\delta\gamma(\delta^2+\gamma^2-8)-2\delta^2\gamma^2-4=0$ is satisfied. Once deautonomised, they correspond to equations (3.14) and (3.8) of [302].

We turn now to the case $\gamma=0$. Clearly (\qtes) now becomes a QRT mapping. Its invariant, given by the limit $\gamma=0$, of (\qpen) obeys  the standard QRT-type conservation relation. While we can now apply the restoration procedure directly, it is preferable to see what happens to the equations for $x$ or $y$ alone. The interesting observation is that $\gamma$ and $\delta$ enter in both (\qhep) and (\pdyo) perfectly symmetrically. Thus we would have found precisely the same equation if instead of taking $\gamma=0$ we had taken $\delta=0$, despite the fact that $\gamma$ and $\delta$ do not play the same role in (\qtes).  (Note that the case $\delta=0$ corresponds precisely to the autonomous limit of case XI we referred to in the introduction).
The canonical form of the invariants is obtained now with $\mu=0$. The corresponding homographic transformations are
$$x_n={1\over 1-X_n},\eqdef\penn$$
and
$$y_n={(\delta+1)^2Y_n+(\delta-1)^2\over Y_n+1}.\eqdef\pdek$$
(Note that in the case $\delta=0$ the first homographic transformation is identical to  (\dqek), provided we invert $X$ and rescale). The resulting equations are of the same form as (\pena) and (\phex) leading again, when deautonomised to equations associated to D$_5^{(1)}$.

Next we consider the restoration towards equations which would have been associated to E$_7^{(1)}$ in the case $\gamma\delta\ne0$ and which are obtained here when either $\gamma=0$ or $\delta=0$. Again we present the final result without presenting all the calculational details. The equation for $X$ is 
$$\left({X_{n+1}X_n-Q_{n+1}Q_n\over X_{n+1}X_n-1}\right)\left({X_nX_{n-1}-Q_nQ_{n-1}\over X_nX_{n-1}-1}\right)={X_n-Z_nZ_{n-1}/R_n\over X_n-1/R_n},\eqdef\sena$$
while the equation for $Y$ is
 $$\left({Y_{n+1}Y_n-Z_{n+1}Z_n\over Y_{n+1}Y_n-1}\right)\left({Y_nY_{n-1}-Z_nZ_{n-1}\over Y_nY_{n-1}-1}\right)={R_{n+1}R_n}{Y_n-Q_{n+1}Q_n\over Y_n-1}.\eqdef\sdyo$$
The Miura relation between the two equations is simply
$$X_n={1\over R_n}\left({Y_nY_{n-1}-Z_nZ_{n-1}\over Y_nY_{n-1}-1}\right)\eqdaf\stri$$
$$Y_n=\left({X_{n+1}X_n-Q_{n+1}Q_n\over X_{n+1}X_n-1}\right).\eqno(\stri b)$$
Equations (\sena) and (\sdyo) were first obtained in [\sequel] where we found the following expressions for the parameters: $\log Q_n=\alpha n+\beta-\alpha/2+2n\phi_2(n)$, $\log R_n=\alpha n/2+\zeta-n\phi_2(n)+\chi_4(n)$, and $\log Z_n=\alpha n+\beta+\chi_4(n)+\chi_4(n+1)$. Both equations, are associated to the affine Weyl group D$_5^{(1)}$ and {\sl not} with E$_7^{(1)}$. So taking $\gamma=0$ or $\delta=0$ does change the space of the discrete Painlev\'e equations resulting from the restoration process. 

\section{ Limiting cases}

In the previous section we have seen that the canonical form for equations associated to D$_5^{(1)}$ is obtained for $\mu_0=\gamma^2\delta^2$ while that for E$_7^{(1)}$-associated equations is obtained for $\mu_{\pm}=(\delta^2+\gamma^2)^2/4-(\delta\pm\gamma)^2$. Let us now suppose that $\gamma$ and $\delta$ are related by $\delta=\gamma$. In this case $\mu_-=\gamma^4$ i.e. precisely the value of $\mu_0$. (Similarly when 
$\delta=-\gamma$, $\mu_+=\mu_0$). It is interesting to see what the result is of the restoration for $\mu_-=\gamma^4$. (This choice of $\delta$ does not affect the restoration with $\mu_+=\gamma^4-4\gamma^2$ which leads again to the same E$_7^{(1)}$-associated equations as before).

As in the previous cases we consider separately the $x$ and the $y$ equation. We start from the former with $\delta=\gamma$ and introduce the transformation
$$x_n={\gamma^3X_n+3\gamma^2-4\over \gamma X_n-1},\eqdef\stes$$
which leads to the mapping
$$X_{n+1}+X_{n-1}={AX_n+B\over X_n^2-1},\eqdef\spen$$
where $A=6/\gamma^2-2$ and $B=-4/\gamma^3$. The deautonomisation of (\spen) leads to $A_n=\alpha(n-1/2)+\beta$ and $B=\zeta+\phi_2(n)$. This is the equation known as the asymmetric discrete P$_{\rm II}$ [\refdef\adptwo] which is associated to the affine Weyl group A$_3^{(1)}$. Similarly starting from the $y$ equation and introducing the transformation
$$y_n={Y_n+8-6/\gamma^2\over Y_n+2/\gamma^2},\eqdef\shex$$
for which we obtain the equation
$$(Y_{n+1}+Y_n)(Y_n+Y_{n-1})={-4Y_n^2+C^2\over Y_n+D},\eqdef\shep$$
where $C=-4/\gamma^3$ and $D=3/\gamma^2-1$. Deautonomising (\shep) we find $D_n=(\alpha n+\beta+\phi_2(n))/2$ and $C=\zeta+\alpha/2$. This is the equation identified in [\refdef\mitfokas] as the discrete analogue of equation 34 in the Painlev\'e-Gambier list. This is true in the ``symmetric'' case, i.e. when the term $\phi_2$ is neglected. Keeping the term  $\phi_2$ we can show that the (\spen) is in fact a discrete analogue of P${\rm III}$, while (\shep) is the discrete analogue of what Ohyama and Okumura [\refdef\oo] call the degenerate P$_{\rm V}$, obtained from P$_{\rm V}$ for $\delta=0$, in Ince's [\refdef\ince] notations. The two equations are related by a Miura transformation, as shown in [\refdef\miuratwo] in the symmetric case. We have indeed the system
$$X_n={Y_n-Y_{n-1}+C\over Y_n+Y_{n-1}}\eqdaf\soct$$
$$Y_n=(X_n+1)(X_{n+1}-1)-D_n,\eqno(\soct b)$$
which extends the result of [\miuratwo] to the case where the $\phi_2(n)$ term is present.

Restoring to `higher' discrete Painlev\'e equations starting from (\spen), (\shep) or (\soct) does not lead to anything new: one finds the same results as in section 3 under the constraint $\delta=\gamma$. The only difference is that here a restoration to D$_5^{(1)}$ associated systems is not possible.

A caveat is in order at this point. If we start from any of the aforementioned equations, say (\spen), and  try to restore towards an equation with E$_6^{(1)}$ symmetry we find that such a restoration is indeed possible. However, upon closer inspection, it turns out that when this is possible we have $A\pm B=0$ and thus (\spen) becomes, after some elementary manipulations 
$$X_{n+1}+X_{n-1}=1+{A\over X_n}.\eqdef\htes$$
The condition $A\pm B=0$ has modified the remnant equation and thus we are not dealing with the same problem any more. 
The deautonomisation of (\htes) leads to $A_n=\alpha n+\beta+\phi_3(n)$ and equation associated to A$_3^{(1)}$, i.e. with the same symmetry group but where a different period has made its appearance. This caveat is not restricted to the specific equations of this section but is actually relevant in general. While attempting a restoration to some family of discrete Painlev\'e equations one must always make sure that the resulting constraints do not modify the remnant equation. 

The final case we shall examine corresponds to $\gamma=\delta=0$. By taking the autonomous limit we obtain the mapping
$${x_{n+1}-4\over x_{n+1}}{x_{n}-4\over x_n}{y_{n}-1\over y_{n}-9}=1,\eqdaf\senn$$
$${1\over y_{n}-1}+{1\over y_{n-1}-1}-{2\over x_{n}-4}=0.\eqno(\senn b)$$
We introduce the homographic transformations
$$X_n={x_n-4\over x_n},\qquad Y_n={y_n-9\over y_n-1},\eqdef\sdek$$
obtaining the remnant system
$$X_{n+1}X_n=Y_n,\eqdaf\hena$$
$$Y_n+Y_{n-1}=A+{B\over X_n},\eqno(\hena b)$$
with $A=6$ and $B=-4$.
Eliminating $X$ or $Y$ from (\hena) is straightforward. We obtain thus the mappings
$$X_{n+1}+X_{n-1}={A\over X_n}+{B\over X_n^2},\eqdef\hdyo$$
and 
$$(Y_{n+1}+Y_n)(Y_n+Y_{n-1})={B^2\over Y_n+A/2},\eqdef\htri$$
where, in (\htri), we have translated $Y_n\to Y_n+A/2$. The deautonomisation of (\hdyo), (\htri) leads to $A_n=\alpha n+\beta$, with $B$ constant, discrete Painlev\'e equations associated to the group A$_1^{(1)}$. Equations (\hdyo) and (\htri) are well-known discrete analogues to Painlev\'e I, and the Miura transformation (\hena) linking them was already obtained in [\mitfokas].

Having the remnant equations (\hdyo), (\htri) or system (\hena) one can implement the restoration procedure in order to obtain, starting from A$_1^{(1)}$ equations associated to ``higher'' groups. However it turns out that the choice is very limited. The only possibility is to use the inverse of the transformations (\sdek) going back to the additive E$_8^{(1)}$-associated equation we have started with (and, of course, its multiplicative and elliptic counterparts). This is one of the rare cases where the deconstruction/restoration approach gives poor results. On the other hand this is, together with a case we have studied in [\ouranton], the only case where starting from an equation in E$_8^{(1)}$  and deautonomising the remnant mapping, we obtain an equation at the other end of the degeneration cascade, namely one associated to A$_1^{(1)}$ (a multiplicative one in that case).

\section{ Conclusions}

In this paper we have presented a further application of the method we introduced in [\resto] and which we have dubbed the deconstruction/restoration procedure. In a nutshell, this method allows one to generate Painlev\'e equations from one given example of such an equation, hopefully leading to new ones (although, given the state of our knowledge on these systems, the derivation of new equations is becoming increasingly infrequent). Typically, in our approach, we start from an equation associated to the affine Weyl group E$_8^{(1)}$. Ideally, we should have worked with the ``highest'' equations of the degeneration cascade, i.e. multiplicative or even elliptic, but we prefer to work with additive ones, since once the latter are known the construction of the multiplicative and elliptic ones is straightforward. Taking the autonomous limit of the discrete Painlev\'e equation we obtain a mapping which we bring to the simplest possible form by a homographic transformation, generating what we called the remnant mapping. Then by applying appropriate homographic transformations we construct all possible integrable mappings, for that starting point, guided by our classification of canonical QRT forms. The deautonomisation of these mappings leads to discrete Painlev\'e equations. However, as shown in this paper, the deconstruction part leading to the remnant mapping is not a mandatory one. One can just as easily work directly with the initial autonomous system and apply the homographic transformations leading to the possible canonical forms to that equation. The usefulness of the remnant system lies merely in the fact that it provides a convenient starting point in the degeneration cascade from which to start the restoration process.

Two discrete Painlev\'e equations, derived in [\deight], were studied in this paper. In the first case we applied the standard deconstruction/restoration approach, obtained the remnant mapping and proceeded from there. Moreover, in this case, since the equation lends itself to such calculations, we also constructed systems corresponding to multistep evolutions. In the second case we proceeded directly to the restoration starting from the initial mapping, without going through the deconstruction phase. This second system, studied in sections 3 and 4 has the feature of leading, at the autonomous limit, to a mapping which is not of QRT type. In this case the restoration procedure cannot be applied, since it is tailored to the QRT case. However, as we have already explained in [\resto], a workaround does exist. It suffices to eliminate either of the two dependent variables and work separately with the mappings  obtained for the remaining variable, which are of QRT type. Once the restoration is performed, one can then in principle derive the Miura transformation between the equations obtained for each variable separatedly, this relation being the result of the restoration on the initial non-QRT system.

A caveat concerning the application of the restoration method was presented in section 4, but its significance is much wider than the simple examples of that section. While applying the restoration method one should always verify that the constraints one obtains do not lead to simplifications (by common factors) in the remnant equation. Whenever such simplifications occur (something we have dubbed ``degeneracy'' [\refdef\capel] in the past) the initial problem is altered, and one would be investigating restorations of a different system. 

An open question does remain, concerning the system, Case X, we studied. While we were able to derive the restoration results for each of the two variables, leading to equations associated to the group E$_7^{(1)}$ we could not, due to the unmanageable bulk of the calculations involved, derive the Miura relating the two equations. What is certain, however, is that such a relation does exist. In fact, the deconstruction/restoration approach helped us realise the ubiquity of such transformations. In future works of ours we intend to address the question of the construction of Miura transformations for discrete Painlev\'e equations in depth. This is doubly interesting, not only because a Miura system relates two discrete Painlev\'e equations, but also because by being an integrable, non-autonomous system, associated to an affine Weyl group, it is also a discrete Painlev\'e equation in its own right.

\section*{References}

\noindent{[\sakai\} H. Sakai, Commun. Math. Phys. 220 (2001) 165.

\noindent{[\ellip]} Y. Ohta, A. Ramani and B. Grammaticos, J. Phys. A 35 (2002) L653.

\noindent{[\eight]} Y. Ohta, A. Ramani and B. Grammaticos, J. Phys. A 34 (2001) 10523.

\noindent{[\multi]} B. Grammaticos and A. Ramani, J. Phys. A 48 (2015) 16FT02.

\noindent{[\jmptri]} B. Grammaticos and A. Ramani, J. Math. Phys. 56 (2015) 083507.

\noindent{[\deight]} A. Ramani and B. Grammaticos, J. Phys. A 48 (2015) 355204.

\noindent{[\resto]} B. Grammaticos, A. Ramani and R. Willox, {\sl Restoring discrete Painlev\'e equations from an E8-associated one}, preprint (2018), arXiv:1812.00712 [math-ph].

\noindent{[\qrt]} G.R.W. Quispel, J.A.G. Roberts and C.J. Thompson, Physica D34 (1989) 183.

\noindent{[\auton]} A. Ramani, S. Carstea, B. Grammaticos and Y. Ohta, Physica A  305 (2002) 437.

\noindent{[\canon]} A. Ramani, B. Grammaticos, J. Satsuma and T. Tamizhmani, J. Phys. A 51 (2018) 395203.

\noindent{[\ouranton]} R. Willox, A. Ramani and B. Grammaticos, J. Math. Phys. 58 (2017) 123504.

\noindent{[\mittami]} A. Ramani, B. Grammaticos and T. Tamizhmani, J. Math. Phys. 59 (2018) 113506.

\noindent{[\qasym]} B. Grammaticos, A. Ramani, K.M. Tamizhmani, T. Tamizhmani and J. Satsuma, J. Math. Phys. 57 (2016) 043506.

\noindent{[\addit]} A. Ramani and B. Grammaticos, J. Phys. A 50 (2017) 055204.

\noindent{[\sequel]} A. Ramani, R. Willox, B. Grammaticos, A.S. Carstea and J. Satsuma, Physica A 347 (2005) 1.

\noindent{[\dps]} A. Ramani, B. Grammaticos and J. Hietarinta, Phys. Rev. Lett. 67 (1991) 1829.

\noindent{[\jimbo]} M. Jimbo and H. Sakai, Lett. Math. Phys. 38 (1996) 145. 

\noindent{[\mitjimbo]} M. Jimbo, H. Sakai, A. Ramani and B. Grammaticos, Phys. Lett. A  217 (1996) 111.

\noindent{[\afb]} A. Ramani and B. Grammaticos, Chaos Sol. Frac. 24 (2005) 1331.

\noindent{[\ohta]} A. Ramani, Y. Ohta and B. Grammaticos, J. Phys. A 39 (2006) 12167.

\noindent{[\adptwo]} A. Ramani, Y. Ohta, J. Satsuma and B. Grammaticos, Comm. Math. Phys. 192 (1998) 67.

\noindent{[\mitfokas]} A. Fokas, B. Grammaticos and A. Ramani, J. of Math. Anal. and Appl. 180 (1993) 342

\noindent{[\oo]} Y.  Ohyama and S. Okumura, J. Phys. A 39 (2006) 12129.

\noindent{[\ince]} E.L. Ince, {\sl Ordinary Differential Equations}, Dover, London, (1956).

\noindent{[\miuratwo]} A. Ramani and B. Grammaticos, Jour. Phys. A 25 (1992) L633.

\noindent{[\capel]} A. Ramani and B. Grammaticos, Physica A 228 (1996) 160.

\end{document}